\documentclass[aps,prd,onecolumn,groupedaddress,showpacs,nofootinbib,amssymb]{revtex4-2}
\usepackage[dvips]{graphicx}
\usepackage{amssymb}
\usepackage{amsmath}
\usepackage{graphicx,color}
\usepackage{amsfonts}
\usepackage{bm}
\usepackage{cancel}
\usepackage{comment}
\usepackage{hyperref}
\usepackage{ulem}

\newcommand{\e}{\mathrm{e}}

\allowdisplaybreaks[4]

\begin{document}

\tolerance=5000

\title{Wormholes inside stars and black holes}
\author{Shin'ichi~Nojiri$^{1,2}$}\email{nojiri@gravity.phys.nagoya-u.ac.jp}
\author{S.D.~Odintsov$^{3,4}$}\email{odintsov@ice.csic.es}
\author{Vladimir~Folomeev$^{5,6}$}\email{vfolomeev@mail.ru}

\affiliation{ 
$^{1)}$ Department of Physics, Nagoya University,
Nagoya 464-8602, Japan \\
$^{2)}$ Kobayashi-Maskawa Institute for the Origin of Particles
and the Universe, Nagoya University, Nagoya 464-8602, Japan \\
$^{3)}$ Institute of Space Sciences (ICE, CSIC) C. Can Magrans
s/n, 08193 Barcelona, Spain \\
$^{4)}$ ICREA, Passeig Luis Companys, 23, 08010 Barcelona, Spain \\
$^{5)}$ Institute of Nuclear Physics, Almaty 050032, Kazakhstan \\
$^{6)}$ Academician J. Jeenbaev Institute of Physics of the NAS of the Kyrgyz Republic,
265 a, Chui Street, Bishkek 720071, Kyrgyzstan 
}

\begin{abstract}
We construct models of two exotic objects: (i)~a wormhole whose throat is hidden by a stellar object like a neutron star; 
and (ii)~a wormhole inside a black hole. 
We work within Einstein's gravity coupled to two scalar fields with a specific choice of the scalar field Lagrangian. 
In general, the model contains ghosts, but they are eliminated using the constraints given by the Lagrange multiplier fields. 
The constraints are a generalization of the mimetic constraint, where non-dynamical dark matter effectively appears. 
As a result, in our model, instead of the non-dynamical dark matter, non-dynamical exotic matter like a phantom effectively arises.  
For the mixed wormhole-plus-star system, we find the corresponding mass-radius relation and show that it is possible to get characteristics comparable to
those of ordinary neutron stars.
For the wormhole inside the black hole, we find an extremal limit where the radius of the throat coincides with the radius of the event horizon and
demonstrate that the Hawking temperature vanishes in this limit. 
\end{abstract}


\maketitle

\section{Introduction}
\label{sec:Int}

Wormholes are exotic compact objects possessing nontrivial spacetime topology and connecting
either different regions of the Universe or separate universes~\cite{Misner:1960zz, Wheeler:1957mu} 
(see also Refs.~\cite{Bronnikov:1973fh, Ellis:1973yv, Morris:1988cz, Morris:1988tu, Visser:1989kh, Visser:1989kg, 
Visser:2003yf} and Ref.~\cite{Visser} for a general overview of Lorentzian wormholes). 
In the framework of Einstein's general relativity, the existence of a static traversable wormhole requires exotic matter which violates the null energy condition 
(at least in the neighbourhood of the wormhole throat~\cite{Hochberg:1997wp}) 
and therefore all the energy conditions~\cite{Poisson:1995sv, Lobo:2005yv, Lobo:2007zb}. 
There are several attempts to construct the traversable wormhole by modifying Einstein's gravity, 
as in the Brans-Dicke theory \cite{Nandi:1997mx, Lobo:2010sb, Sushkov:2011zh} and other theories~\cite{Bronnikov:2010tt}. 

In this paper, we consider more exotic objects, that is, a wormhole whose throat is hidden by a stellar object like a neutron star 
and a wormhole inside a black hole. 
The wormhole inside a stellar object has been studied in 
Refs.~\cite{Dzhunushaliev:2011xx,Dzhunushaliev:2012ke,Dzhunushaliev:2013lna,Dzhunushaliev:2014mza,Aringazin:2014rva,Dzhunushaliev:2015sla,Dzhunushaliev:2016ylj,Dzhunushaliev:2022elv}
numerically within Einstein's gravity. 
More general geometry of the wormhole inside a black hole has been well-investigated in Ref.~\cite{Simpson:2018tsi}. 
Such mixed neutron-star-plus-wormhole configurations containing a ghost scalar
field (which ensures the presence of a nontrivial spacetime
topology) and ordinary neutron star matter possess properties both of wormholes and of
usual stars: their masses and sizes are comparable to those of typical ordinary neutron stars. 
However, the presence of a ghost field results in distressing consequences: both the 
mixed configurations and pure wormholes are dynamically unstable objects~\cite{Shinkai:2002gv, Gonzalez:2008wd,  Gonzalez:2008xk, Bronnikov:2011if, Dzhunushaliev:2013lna}.
Such instability is caused essentially by the presence in the system of ghost fields. 
This motivates one to study the possibilities of providing a nontrivial spacetime topology without using ghost fields. 

The aforementioned mixed neutron-star-plus-wormhole configurations have been constructed by solving self-consistently the Einstein equations
and equations for a scalar field and neutron matter. 
In doing so, as in the case of ordinary neutron stars, some characteristic mass-radius relations are obtained,
whose form strongly depends on a specific choice of an equation of state of neutron matter and the type of the scalar field.
In the present paper, we will use another approach when the gravitational field of a system is specified by giving some \textit{ad hoc}
distribution of an energy density of neutron matter threading the wormhole, as it is also done, for example, in 
Refs.~\cite{Sushkov:2005kj,Dymnikova:2003vt,Lobo:2012qq}. 
In this case, the mass-radius relations will already differ substantially from the dependencies typical of ordinary neutron stars or 
the mixed neutron-star-plus-wormhole configurations mentioned above.
In particular, for the neutron-matter energy-density profile used in the present paper, 
the mass of pure stars (i.e., configurations without a wormhole)  turns out to be proportional to the third power of the radius, 
whereas the ADM mass of the wormhole whose throat is hidden by a stellar object under consideration is proportional to the radius of the stellar object, 
as in the case of the Schwarzschild black hole. 
In turn, the properties of the wormhole inside the black hole are very similar to those of the Reissner-Nordstr\"{o}m black hole, although there is only one event horizon. 
For example, there appears an extremal limit where the radius of the throat coincides with the radius of the event horizon. 
In this limit, we find that the Hawking temperature vanishes. 

The key idea for the construction of the exotic objects in the present paper is that we employ the formulation 
by using two scalar fields suggested in Ref.~\cite{Nojiri:2020blr}, where it has been 
shown that {an} arbitrarily given spherically symmetric and static/time-dependent geometry can be realized 
in the framework of Einstein's gravity coupled with two scalar fields. 
However, in the model of Ref.~\cite{Nojiri:2020blr}, there appear ghosts, and this means that the model is inconsistent. 
In classical theories, the kinetic energy of the ghost is unbounded below. 
In the framework of quantum theories, on the other hand, 
the ghosts generate negative norm states, which break the Copenhagen interpretation of the quantum theories~\cite{Kugo:1979gm}. 
The ghosts can be, however, eliminated by using constraints given by the Lagrange multiplier fields, as shown in Refs.~\cite{Nojiri:2023dvf, Nojiri:2023zlp, Elizalde:2023rds, Nojiri:2023ztz}. 
These constraints can be regarded as a generalization of the mimetic constraint of Ref.~\cite{Chamseddine:2013kea}, where non-dynamical dark matter effectively appears. 
In our model, non-dynamical exotic matter like a phantom effectively appears as well. 

The paper is organized as follows. 
In Sec.~\ref{sec:Model},  we review the formulation of Einstein's gravity coupled with two scalar fields and give the explicit forms of the 
Einstein equations and the field equations for the two scalar fields. 
Furthermore, we explain how the ghosts can be eliminated and how arbitrarily given spherically symmetric and static/time-dependent geometries are realized. 
In Sec.~\ref{WormholeinStellarOject}, by using the formulation of Sec.~\ref{sec:Model}, we construct a model realizing a wormhole 
where a stellar object like a neutron star could hide the throat. 
Also, here we discuss the energy conditions and compare the mixed object obtained with ordinary neutron stars. 
In Sec.~\ref{WormholeinBlackHole}, we consider a wormhole inside the event horizon of a black hole. 
The wormhole could connect two universes by the throat, but if the black hole in our universe is really a black hole, 
the object in another universe connected by the throat can be regarded as a white hole. 
Finally, Sec.~\ref{SummaryDiscussion} is devoted to summary and discussion. 

\section{Einstein's gravity coupled to two scalars}
\label{sec:Model}

The arbitrarily given spherically symmetric spacetimes can be realized in the framework of Einstein's gravity coupled with two scalar fields 
even if the spacetime is dynamical or time-dependent. 
In the model of Ref.~\cite{Nojiri:2020blr}, there appear ghosts, which make the model inconsistent because the kinetic energy of the ghost is unbounded below 
in the framework of {\sout{the}} classical mechanics and the ghosts generate negative norm states, which break the Copenhagen interpretation 
of quantum theories \cite{Kugo:1979gm}. 
In the subsequent works~\cite{Nojiri:2023dvf, Nojiri:2023zlp, Elizalde:2023rds, Nojiri:2023ztz}, it has been shown that the ghosts can be eliminated 
by constraints given by the Lagrange multiplier fields. 
These constraints are similar to the mimetic constraint of Ref.~\cite{Chamseddine:2013kea}, where non-dynamical dark matter effectively appears. 

The action in the model of Ref.~\cite{Nojiri:2020blr} is given by that of four-dimensional Einstein's gravity coupled with two scalar fields $\phi$ and $\chi$,
\begin{align}
\label{I8}
S_{\mathrm{GR} \phi\chi} = \int d^4 x \sqrt{-g} &\, \left[ \frac{R}{2\kappa^2}
 - \frac{1}{2} A (\phi,\chi) \partial_\mu \phi \partial^\mu \phi
 - B (\phi,\chi) \partial_\mu \phi \partial^\mu \chi \right. \nonumber \\
& \left. \qquad - \frac{1}{2} C (\phi,\chi) \partial_\mu \chi \partial^\mu \chi
 - V (\phi,\chi) + \mathcal{L}_\mathrm{matter} \right]\, .
\end{align}
Here $A(\phi,\chi)$, $B(\phi,\chi)$, and $C(\phi,\chi)$ are arbitrary functions, $V(\phi,\chi)$ is the scalar-field potential, 
and $\mathcal{L}_\mathrm{matter}$ is the matter Lagrangian density. 
The gravitational coupling constant $\kappa$ is defined by using Newton's gravitational constant $G$, $\kappa^2=8\pi G$, 
although we mainly use geometrized units $c=G=1$ throughout the paper. 

By the variation of the action \eqref{I8} with respect to the metric $g_{\mu\nu}$, we obtain the Einstein equations, 
\begin{align}
\label{gb4bD4}
0= &\, \frac{1}{2\kappa^2}\left(- R_{\mu\nu} + \frac{1}{2} g_{\mu\nu} R\right) \nonumber \\
&\, + \frac{1}{2} g_{\mu\nu} \left[
 - \frac{1}{2} A (\phi,\chi) \partial_\rho \phi \partial^\rho \phi
 - B (\phi,\chi) \partial_\rho \phi \partial^\rho \chi
 - \frac{1}{2} C (\phi,\chi) \partial_\rho \chi \partial^\rho \chi - V (\phi,\chi)\right] \nonumber \\
&\, + \frac{1}{2} \left[ A (\phi,\chi) \partial_\mu \phi \partial_\nu \phi
+ B (\phi,\chi) \left( \partial_\mu \phi \partial_\nu \chi
+ \partial_\nu \phi \partial_\mu \chi \right)
+ C (\phi,\chi) \partial_\mu \chi \partial_\nu \chi \right] 
+ \frac{1}{2} T_{\mathrm{matter}\, \mu\nu} \, ,
\end{align}
where the Greek indices run over $\mu, \nu, \ldots=0, 1, 2, 3$ 
and $T_{\mathrm{matter}\, \mu\nu}$ is the energy-momentum tensor of matter. 
On the other hand, the variations of the action \eqref{I8} with respect to the fields $\phi$ and $\chi$ give the following equations: 
\begin{align}
\label{I10}
0 =&\, \frac{1}{2} A_\phi \partial_\mu \phi \partial^\mu \phi
+ A \nabla^\mu \partial_\mu \phi + A_\chi \partial_\mu \phi \partial^\mu \chi
+ \left( B_\chi - \frac{1}{2} C_\phi \right)\partial_\mu \chi \partial^\mu \chi
+ B \nabla^\mu \partial_\mu \chi - V_\phi \, , \nonumber \\
0 =&\, \left( - \frac{1}{2} A_\chi + B_\phi \right) \partial_\mu \phi \partial^\mu \phi
+ B \nabla^\mu \partial_\mu \phi
+ \frac{1}{2} C_\chi \partial_\mu \chi \partial^\mu \chi + C \nabla^\mu \partial_\mu \chi
+ C_\phi \partial_\mu \phi \partial^\mu \chi - V_\chi \, ,
\end{align}
where $A_\phi=\partial A(\phi,\chi)/\partial \phi$, etc. 
These field equations are nothing but the Bianchi identities. 
In Appendix~\ref{Bianchi}, we show that the field equations (\ref{I10}) can be surely obtained from the Einstein equations~(\ref{gb4bD4}).

\subsection{Elimination of the ghosts}
\label{sec:Elimitation}

The metric of a general spherically symmetric and time-dependent spacetime is given by,
\begin{align}
\label{GBiv_time}
ds^2 = - \e^{2\nu (t,r)} dt^2 + \e^{2\lambda (t,r)} dr^2 + r^2 \left( d\vartheta^2 + \sin^2\vartheta d\varphi^2 \right)\, .
\end{align}
The argument that the metric in (\ref{GBiv_time}) is a general one is given in Ref.~\cite{Nojiri:2020blr} (see also the textbook~\cite{LL}). 
We may also assume the ansatz 
\begin{align}
\label{TSBH1}
\phi=t\, , \quad \chi=r\, ,
\end{align}
which does not give any loss of generality. 
The detailed arguments have been given in the previous papers~\cite{Nojiri:2020blr, Nojiri:2023dvf, Nojiri:2023zlp, Elizalde:2023rds, Nojiri:2023ztz}. 

In Sec.~\ref{WormholeinStellarOject}, we discuss how we can construct models realizing an arbitrarily given spherically
symmetric geometry expressed by the metric (\ref{GBiv_time}). 
In the realizations, however, the functions $A$ and/or $C$ are often negative, and therefore $\phi$ and/or $\chi$ become ghosts. 
In order to eliminate the ghosts, we impose constraints by introducing the Lagrange multiplier fields $\lambda_\phi$ and $\lambda_\chi$ 
and modifying the action (\ref{I8}) $S_{\mathrm{GR} \phi\chi} \to S_{\mathrm{GR} \phi\chi} + S_\lambda$, 
where the additional term $S_\lambda$ is given by
\begin{align}
\label{lambda1}
S_\lambda = \int d^4 x \sqrt{-g} \left[ \lambda_\phi \left( \e^{-2\nu(t=\phi, r=\chi)} \partial_\mu \phi \partial^\mu \phi + 1 \right)
+ \lambda_\chi \left( \e^{-2\lambda(t=\phi, r=\chi)} \partial_\mu \chi \partial^\mu \chi - 1 \right) \right] \, .
\end{align}
By varying $S_\lambda$ with respect to $\lambda_\phi$ and $\lambda_\chi$, we obtain the following constraints:
\begin{align}
\label{lambda2}
0 = \e^{-2\nu(t=\phi, r=\chi)} \partial_\mu \phi \partial^\mu \phi + 1 \, , \quad
0 = \e^{-2\lambda(t=\phi, r=\chi)} \partial_\mu \chi \partial^\mu \chi - 1 \, ,
\end{align}
which is consistent with the assumption (\ref{TSBH1}).
The constraints (\ref{lambda2}) are similar to the mimetic constraint of Ref.~\cite{Chamseddine:2013kea}, where non-dynamical dark matter appears. 

The constraints from Eq.~(\ref{lambda2}) make the scalar fields $\phi$ and $\chi$ non-dynamical, and the fluctuations of
$\phi$ and $\chi$ around the background (\ref{TSBH1}) do not propagate. 
We now write the fluctuations as follows: 
\begin{align}
\label{pert1}
\phi=t + \delta \phi \, , \quad \chi=r + \delta \chi\, ,
\end{align}
and, by using Eq.~(\ref{lambda2}), we find 
\begin{align}
\label{pert2}
\partial_t \left( \e^{-2\nu(t,r)} \delta \phi \right) = \partial_r \left( \e^{-2\lambda(t,r)} \delta \chi \right) = 0\, .
\end{align}
The equation~(\ref{pert2}) tells us that, by imposing the initial condition $\delta\phi=0$ and by imposing the boundary condition $\delta\chi\to 0$ when $r\to \infty$, 
we can find that both of $\delta \phi$ and $\delta \chi$ vanish in the whole spacetime, $\delta\phi=0$ and $\delta\chi=0$. 
This tells us that both $\phi$ and $\chi$ are non-dynamical or frozen degrees of freedom. 

As partially or completely shown in Refs.~\cite{Nojiri:2023dvf, Nojiri:2023zlp, Elizalde:2023rds, Nojiri:2023ztz}, 
even in the model given by the modified action $S_{\mathrm{GR} \phi\chi} + S_\lambda$, $\lambda_\phi=\lambda_\chi=0$ 
consistently appear as a solution and therefore any solution of Eqs.~(\ref{gb4bD4}) and (\ref{I10}) which are based on the original action (\ref{I8}) 
is a solution even for the modified model with the action $S_{\mathrm{GR} \phi\chi} + S_\lambda$. 

\subsection{Reconstruction of models which realize any given spherically 
symmetric and static/time-dependent spacetime} 
\label{sec:Reconstruction}

We now try to construct a model which has a solution realizing the metric functions $\e^{2\nu(t,r)}$ and $\e^{2\lambda(t,r)}$ given in Eq.~(\ref{GBiv_time}). 

The $(t,t)$, $(r,r)$, $(\vartheta,\vartheta)$, and $(t,r)$ components of Eqs.~(\ref{gb4bD4}) have the following form:
\begin{align}
\label{TSBH2}
\frac{\e^{-2\lambda + 2\nu}}{\kappa^2} \left( \frac{2\lambda'}{r} + \frac{\e^{2\lambda} - 1}{r^2} \right)
=&\, - \e^{2\nu} \left( - \frac{A}{2} \e^{-2\nu} - \frac{C}{2} \e^{-2\lambda} - V \right) + \e^{2\nu} \rho \, , \nonumber \\
\frac{1}{\kappa^2} \left( \frac{2\nu'}{r} - \frac{\e^{2\lambda} - 1}{r^2} \right)
=&\, \e^{2\lambda} \left( \frac{A}{2} \e^{-2\nu} + \frac{C}{2} \e^{-2\lambda} - V \right) + \e^{2\lambda} p \, , \nonumber \\
\frac{1}{\kappa^2} \left\{ - r^2 \e^{-2 \nu} \left[ \ddot\lambda \right. \right. & + \left. \left. \left( \dot\lambda - \dot\nu \right) \dot\lambda \right]
+ \e^{-2\lambda}\left( r \left(\nu' - \lambda' \right) + r^2 \nu'' + r^2 \left( \nu' - \lambda' \right) \nu' \right) \right\} \nonumber \\
=&\, r^2 \left( \frac{A}{2} \e^{-2\nu} - \frac{C}{2} \e^{-2\lambda} - V \right) + r^2 p \, , \nonumber \\
\frac{2\dot\lambda}{\kappa^2 r} =&\, B \, ,
\end{align}
where the dot and prime denote differentiation with respect to the time coordinate $t$ and radial coordinate $r$, respectively. 
In this paper, we assume that the matter is a perfect fluid with $\rho$ and $p$ being the energy density and the pressure of matter, defined by
\begin{align}
\label{FRk2}
T_{\mathrm{matter}\, tt} =-g_{tt}\rho\ ,\quad T_{\mathrm{matter}\, ij}=p\, g_{ij}\, ,
\end{align}
{where $i, j= r,\vartheta, \varphi$.}
The equations (\ref{TSBH2}) can be algebraically solved with respect to $A$, $B$, $C$, and $V$ as follows:
\begin{align}
\label{ABCV}
A=& \frac{\e^{2\nu}}{\kappa^2} \left\{ - \e^{-2 \nu} \left[ \ddot\lambda + \left( \dot\lambda - \dot\nu \right) \dot\lambda \right]
+ \e^{-2\lambda}\left[ \frac{\nu' + \lambda'}{r} + \nu'' + \left( \nu' - \lambda' \right) \nu' + \frac{\e^{2\lambda} - 1}{r^2}\right] \right\}
 - \e^{2\nu} \left( \rho + p \right) \, , \nonumber \\
B=&\, \frac{2\dot\lambda}{\kappa^2 r} \, , \nonumber \\
C=&\, \frac{\e^{2\lambda}}{\kappa^2} \left\{ \e^{-2 \nu} \left[ \ddot\lambda + \left( \dot\lambda - \dot\nu \right) \dot\lambda \right]
 - \e^{-2\lambda}\left[ - \frac{\nu' + \lambda'}{r} + \nu'' + \left( \nu' - \lambda' \right) \nu' + \frac{\e^{2\lambda} - 1}{r^2}\right] \right\} 
\, , \nonumber \\
V=& \frac{\e^{-2\lambda}}{\kappa^2} \left( \frac{\lambda' - \nu'}{r} + \frac{\e^{2\lambda} - 1}{r^2} \right) - \frac{1}{2} \left( \rho - p \right) \, .
\end{align}
Then we can obtain a model that realizes the spacetime defined by the metric (\ref{GBiv_time}) by finding $(t,r)$-dependence of $\rho$ and $p$ and 
by replacing $(t,r)$ in Eq.~(\ref{ABCV}) with $(\phi,\chi)$.

In the following, by using the above model of the two scalar fields, 
we construct a model whose solution is a wormhole inside a stellar object. 
In general, the fluid which forms the wormhole generates instability because the fluid violates the energy conditions. 
Due to the constraints (\ref{lambda2}), the two scalar fields do not propagate, nor do they fluctuate and are therefore non-dynamical. 
This tells us that in the wormhole solution given by the two scalar fields, there does not appear any instability, which is different from the 
standard but exotic fluid which is often used to construct wormhole solutions.

\section{Wormhole inside a stellar object}
\label{WormholeinStellarOject}

We may consider a wormhole whose throat is hidden by a stellar object like a neutron star. 
This can be done by choosing the radius of the neutron star to be larger than the radius of the throat. 
Similar objects have been numerically studied in 
Refs.~\cite{Dzhunushaliev:2011xx, Dzhunushaliev:2012ke, Dzhunushaliev:2013lna, Dzhunushaliev:2014mza, Aringazin:2014rva, 
Dzhunushaliev:2015sla, Dzhunushaliev:2016ylj, Dzhunushaliev:2022elv}, where ghost scalar fields are used to realize the objects. 
In the present paper, we construct the object analytically, and the ghost is eliminated by using the constraint of Sec.~\ref{sec:Elimitation}. 

To this end, a general spherically symmetric and static spacetime whose metric can be obtained from Eq.~\eqref{GBiv_time} in the form, 
\begin{align}
\label{GBiv}
ds^2 = - \e^{2\nu (r)} dt^2 + \e^{2\lambda (r)} dr^2 + r^2 \left( d\vartheta^2 + \sin^2\vartheta d\varphi^2 \right)\, .
\end{align}
In the case of a wormhole whose throat has a radius $r_0$, near the throat $r\sim r_0$, the metric functions $\e^{2\nu}$ and $\e^{2\lambda}$ behave as 
\begin{align}
\label{throat1}
\e^{2\nu}\sim \e^{2\nu_0} + \nu_1 \left(r-r_0\right) \, , \quad 
\e^{2\lambda}\sim \frac{r_0 }{r- r_0}\e^{2\lambda_0}\, ,
\end{align}
with constants $\nu_0$, $\nu_1$, and $\lambda_0$. 
If we redefine the radial coordinate $r$ by using 
\begin{align}
\label{l}
l = 2 \sqrt{r_0 \left(r - r_0\right)}\, ,
\end{align}
near the throat, the metric (\ref{GBiv}) takes the form
\begin{align}
\label{whmetric1}
ds^2 \sim - \left( \e^{2\nu_0} + \frac{\nu_1 l^2}{4r_0} \right) dt^2 + \e^{2\lambda_0}dl^2 
+ {r_0}^2 \left( d\vartheta^2 + \sin^2\vartheta d\varphi^2 \right)\, .
\end{align}
Notice that in Eq.~(\ref{l}) the radial coordinate $l$ is defined to be positive, but in the expression (\ref{whmetric1}), 
we may analytically continue $l$ into the region where $l$ is negative. 
Let us assume that the region where $l$ is positive corresponds to our universe, and then the region where $l$ is negative corresponds to another universe 
connected with our universe by the wormhole. 

We now consider a stellar-like object similar to a neutron star supported by a perfect fluid. 
The conservation law of the perfect fluid with the energy density $\rho$ and the pressure $p$ is given by, 
\begin{equation}
\label{FRN2}
0 = \nabla^\mu T_{\mu r} =\nu' \left( \rho + p \right) + p' \, ,
\end{equation}
while other components of the conservation law are trivially satisfied.
If the equation of state (EoS) $\rho=\rho(p)$ is given, Eq.~(\ref{FRN2}) can be integrated as
\begin{equation}
\label{FRN3}
\nu = - \int^{p(r)}\frac{dp}{\rho(p) + p} \, .
\end{equation}

The EoS of compact stars, like neutron stars, can
be approximated by a polytropic EoS that gives a more or less realistic description of neutron matter at high densities. 
Here we consider two simple EoSs: 
\begin{enumerate}
\item The rest-mass-polytrope,
\begin{equation}
\label{polytrope}
p_1 = K_1 \rho_\mathrm{b}^{1 + \frac{1}{n_1}}\,.
\end{equation}
Here $\rho_\mathrm{b}=n_\mathrm{b} m_\mathrm{b}$ is the rest-mass density of the neutron fluid,
where  $n_\mathrm{b}$ is the baryon number density and $m_\mathrm{b}$ is the baryon mass.
The values of the constants $K_1$ and $n_1$ depend on the properties of the fluid under consideration. 
Such an EOS was used, for instance, in Refs.~\cite{Tooper:1964, Horvat:2010xf, Herrera:2013fja, Yakov} in modeling general relativistic
isotropic and anisotropic fluid spheres. 
In what follows, we refer to this choice as EoS1.
It is known that for the neutron stars, $n_1$ can take values in
the range $0.5\leq n_1 \leq 3$ \cite{Yakov}.
\item  The parametric relation between the pressure and energy density,
\begin{equation}
\label{MassPolytropicEOS}
\rho = \rho_\mathrm{b} + n_2 p_2 \, , \quad p_2 = K_2 \rho_\mathrm{b}^{1+\frac{1}{n_{2}}} \, ,
\end{equation}
with the constant $K_2=k c^2 (n_\mathrm{b}^{(\mathrm{ch})} m_\mathrm{b})^{-1/n_2}$, where $c$ is the velocity of light, 
$n_\mathrm{b}^{(\mathrm{ch})}$ is a characteristic value of $n_\mathrm{b}$, 
and $k$ and $n_2$ are parameters whose values depend on the properties of the neutron matter (see Sec.~\ref{comp_NS_mix}).
In what follows we refer to this choice as EoS2. 
\end{enumerate}

Upon substituting the EoSs \eqref{polytrope} and \eqref{MassPolytropicEOS} into the conservation law \eqref{FRN3},
one can find the following expressions for the metric function $\nu$:
\begin{eqnarray}
\label{nu_1}
&&\text{for the EoS1}: \quad \nu=\nu_c -(n_1+1)\ln{\left(1+K_1 \rho_\mathrm{b}^{1/n_1}\right)} ;\\
&&\text{for the EoS2}: \quad \nu=\nu_c -\ln{\left[1+K_2 (n_2+1) \rho_\mathrm{b}^{1/n_2}\right]} ,
\label{nu_2}
\end{eqnarray}
where $\nu_c$ is an integration constant.

Given one of the aforementioned EoSs, let us consider an example case 
where we assume the following profiles for
$\rho = \rho(r)$ and $\lambda = \lambda(r)$: 
\begin{align}
\label{anz1WH}
\rho=\left\{ \begin{array}{cc}
\rho_c \left[ 1 - \frac{\left(r - r_0\right) ^2}{{\left(R_s - r_0\right)}^2} \right] & \ \mbox{when}\ r_0\leq r\leq R_s \\
0 & \  \mbox{when}\ r>R_s
\end{array} \right. \, , \quad
\e^{-2\lambda} = 1 - \frac{r_0}{r} \, ,
\end{align}
where $r_0$ is a constant corresponding to the radius of the throat, $\rho_c$ is the value of the energy density at the throat $r=r_0$, 
and $R_s$ is the radius of the surface of the neutron fluid. 

For both polytropes \eqref{polytrope} and \eqref{MassPolytropicEOS}, consider the case $n_1=n_2=1$. 
Then, by substituting the expression \eqref{anz1WH} into \eqref{nu_1} and \eqref{nu_2}, we find in the region $r_0\leq r\leq R_s$:
\begin{align}
\label{nuinside1}
&\text{for the EoS1}: \quad \e^{2\nu}= \frac{\e^{2\nu_c}}{\left[ 1 + K_1 \rho_c \left( 1 - \frac{\left(r - r_0\right) ^2}{{\left(R_s - r_0\right)}^2} \right) \right]^4}\, ,
\quad \left( \e^{2\nu}\right)'= \frac{ 8K_1 \rho_c \e^{2\nu_c}\left(r - r_0\right)\left(R_s - r_0\right)^{-2}}
{\left[ 1 + K_1 \rho_c \left( 1 - \frac{\left(r - r_0\right) ^2}{\left(R_s - r_0\right)^2} \right) \right]^5}\, ; \\
&\text{for the EoS2}: \quad \e^{2\nu}= \frac{\e^{2\nu_c}}{\left[ 1 + 2 K_2 \rho_c \left( 1 - \frac{\left(r - r_0\right) ^2}{{\left(R_s - r_0\right)}^2} \right) \right]^2} ,
\quad \left( \e^{2\nu}\right)'= \frac{ 8K_2 \rho_c \e^{2\nu_c}\left(r - r_0\right)\left(R_s - r_0\right)^{-2}}
{\left[ 1 + 2 K_2 \rho_c \left( 1 - \frac{\left(r - r_0\right) ^2}{\left(R_s - r_0\right)^2} \right) \right]^3}\, .
\label{nuinside2}
\end{align}
Then at the surface $r=R_s$, we find for both EoSs
\begin{align}
\label{nuinside3}
\e^{2\nu\left(r=R_s\right)}= \e^{2\nu_c} \, , \quad 
\left. \left( \e^{2\nu}\right)' \right|_{r=R_s}= \frac{8K_{1,2} \rho_c }{R_s - r_0} \e^{2\nu_c}\, .
\end{align}

Next, as an example, we may assume that outside the fluid, i.e., when $r>R_s$, the metric function
\begin{equation}
\label{nu_outside}
\e^{2\nu}=\e^{-2\lambda}=1 - \frac{r_0}{r} . 
\end{equation}
This implies that we deal with asymptotically flat spacetime. 
In turn, the continuities of $\e^{2\nu}$ and $\left(\e^{2\nu}\right)'$ at the surface of the fluid give 
\begin{align}
\label{nuinside4}
\e^{2\nu_c} = 1 - \frac{r_0}{R_s}\, , \quad 8K_{1,2} \rho_c R_s = r_0\, .
\end{align}
These expressions determine the size of the throat $r_0$ and the value of the integration constant $\nu_c$ in terms of $\rho_c$ and $R_s$.
Note here that in the case of $n_{1,2}\neq 1$ the ansatz \eqref{nu_outside} results in $r_0=0$. 

The ansatz \eqref{nu_outside} for the metric functions outside the fluid implies 
that the ADM mass $M$ is given by 
\begin{align}
\label{ADMmass}
M=\frac{r_0}{2}= 4 K_{1,2} \rho_c R_s \, .
\end{align}
This yields the mass-radius relation of the configuration under consideration. 
In Sec.~\ref{comp_NS_mix}, we will compare this mass-radius relation with that of 
ordinary neutron stars described by the same EoSs. 

But before note that, 
in ordinary stars modeled by the toy profiles of the type~\eqref{anz1WH}, 
the mass could be proportional to the third power of the radius $M\propto {R_s}^3$ if the matter is not so compressed 
because the mass $M$ could be proportional to the volume $\sim \frac{4\pi}{3}{R_s}^3$. 
For example, instead of (\ref{anz1WH}), we may consider the following matter profile 
 in a compact star without a wormhole $\left( r_0 \to 0\right)$: 
\begin{align}
\label{anz1}
\rho=\left\{ \begin{array}{cc}
\rho_c \left( 1 - \frac{r^2}{R_s^2} \right) & \ \mbox{when}\ r\leq R_s \\
0 & \ \ \mbox{when}\ r>R_s
\end{array} \right. \, .
\end{align}
Then the mass $M_\mathrm{\star}$ corresponding to
ordinary matter of the compact star, which is generally different from the ADM mass, is given by 
\begin{align}
\label{MRs}
M_\mathrm{\star} =4\pi  \int_0^{R_s} dr\, r^{2} \rho(r) = 4\pi \rho_c \int_0^{R_s} dr\, r^2 \left( 1 - \frac{r^2}{R_s^2} \right)
= \frac{8\pi \rho_c R_s^3}{15} \, ,
\end{align}
and we really find that $M_\mathrm{\star} \propto {R_s}^3$. 
Note that this mass-radius relation differs strongly from that typical of realistic neutron stars (see, e.g., Refs.~\cite{Horvat:2010xf, Herrera:2013fja, Yakov, Ozel:2010fw}).
The reason is that we employ the toy profile \eqref{anz1} which enables us to find the analytical expression~\eqref{MRs}, in contrast to the realistic neutron stars 
whose models are constructed only numerically.

In the case of the Schwarzschild black hole, the ADM mass $M$ is proportional to the Schwarzschild radius. 
If the energy density $\rho_c$ at the throat has a maximum, as it takes place for ordinary neutron stars, 
Eq.~(\ref{ADMmass}) tells us that the mass $M$ is proportional to the radius $R_s$ of the stellar-like object, as in the case of the Schwarzschild black hole,
although $R_s$ must be larger than the Schwarzschild radius $r_0$. 

\subsection{Energy conditions}

Consider now the null, weak, strong, and dominant energy conditions, which state that 
the energy-momentum tensor $T_{\mu\nu}$ satisfies, respectively, the inequalities 
\begin{align}
&T_{\mu\nu} k^\mu k^\nu \geq 0\, , \quad T_{\mu\nu} V^\mu V^\nu \geq 0\, , \quad
\left(T_{\mu\nu}-\frac{1}{2}g_{\mu\nu}T \right)V^\mu V^\nu \geq 0\, , \nonumber \\
&T_{\mu\nu} V^\mu V^\nu \geq 0\, , \quad \text{and} \quad T_{\mu\nu} V^\nu \quad \text{is not spacelike} \, .\nonumber
\end{align}
for any null vector $k^\mu$, $g_{\mu\nu}k^\mu k^\nu =0$, and for any timelike vector $V^\mu$, $g_{\mu\nu}V^\mu V^\nu <0$ (see, e.g., Ref.~\cite{Visser}). 
For the system under consideration, we have the energy-momentum tensor 
$T^\mu_{\ \nu}=\text{diag}\left(-\rho_\mathrm{tot}, p_\mathrm{tot}^r, p_\mathrm{tot}^\vartheta, p_\mathrm{tot}^\vartheta\right)$,
where the subscript ``tot'' denotes the total (the scalar fields plus the fluid) energy density and pressure.
In terms of these quantities, the above conditions yield
\begin{eqnarray}
\text{NEC}:&& \quad \rho_\mathrm{tot}+ p_\mathrm{tot}^r \geq 0\, , \quad \rho_\mathrm{tot}+ p_\mathrm{tot}^\vartheta \geq 0\, , \nonumber \\
\text{WEC}:&& \quad \rho_\mathrm{tot}+ p_\mathrm{tot}^r \geq 0\, , \quad \rho_\mathrm{tot}+ p_\mathrm{tot}^\vartheta \geq 0\, , \quad \rho_\mathrm{tot}\geq 0\, ,\nonumber \\
\text{SEC}:&& \quad \rho_\mathrm{tot}+ p_\mathrm{tot}^r \geq 0\, , \quad \rho_\mathrm{tot}+ p_\mathrm{tot}^\vartheta \geq 0\, , \quad \rho_\mathrm{tot}+p_\mathrm{tot}^r
+2 p_\mathrm{tot}^\vartheta\geq 0\, ,\nonumber \\
\text{DEC}:&& \quad \rho_\mathrm{tot}\geq 0\, , \quad  \rho_\mathrm{tot}\geq | p_\mathrm{tot}^r|\, , \quad  \rho_\mathrm{tot}\geq | p_\mathrm{tot}^\vartheta| \, .\nonumber
\end{eqnarray}

With the ansatz for $\lambda$ in the form \eqref{anz1WH} and \eqref{nu_outside}, we have the following expressions for the principal pressures valid 
over the whole space (inside and outside the fluid):
\begin{align}
\label{rho_tot}
\rho_\mathrm{tot}=&\, 0, \\
p_\mathrm{tot}^r =&\, \frac{2 r(r-r_0)\nu^\prime-r_0}{\kappa^2 r^3},\\
p_\mathrm{tot}^\vartheta =&\, \frac{2 r^2(r-r_0)\left(\nu^{\prime\prime}+\nu^{\prime 2}\right)+r(2 r-r_0)\nu^{\prime}+r_0}{2\kappa^2 r^3}\, .
\end{align}
Using these expressions, one can find in the limit $r\to r_0$
\[
\left(\rho_\mathrm{tot}+ p_\mathrm{tot}^r\right) \to -\frac{1}{\kappa^2 r_0^2}\, , \quad \left(\rho_\mathrm{tot}+ p_\mathrm{tot}^\vartheta\right) \to \frac{1+r_0\nu^\prime(r_0)}{2\kappa^2 r_0^2}\, .
\]
For the choice of the metric function $\nu$ in the form \eqref{nuinside1} and \eqref{nuinside2}, $\nu^\prime(r_0)=0$. 
Then it is evident from these expressions that all the above energy conditions are violated in the vicinity of the throat.

The violation of the energy conditions usually generates the instability of the configuration. 
For example, the speed of sound often becomes larger than the speed of light. 
If we construct a wormhole spacetime by using the two-scalar model in the last section, however, any instability does not appear because 
the two scalar fields are non-dynamical and they do not propagate nor fluctuate, and therefore there does not appear sound generated by the oscillation 
of the effective fluid created by the two scalar fields, although the effective fluid violates the energy conditions.

\subsection{Comparison with ordinary neutron stars}
\label{comp_NS_mix}

\begin{figure}[t]
    \begin{minipage}[t]{.49\linewidth}
        \begin{center}
\includegraphics[width=1.\linewidth]{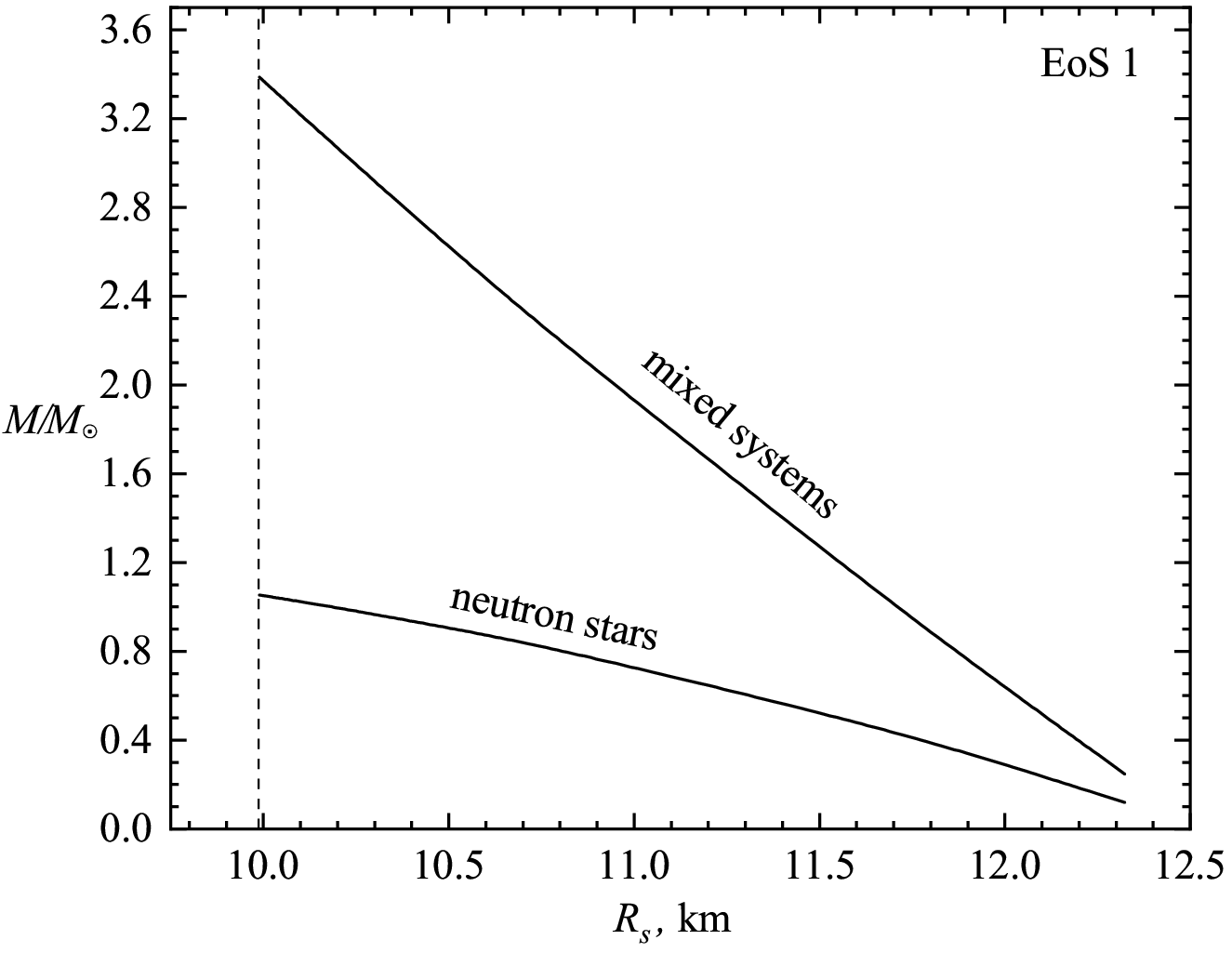}
        \end{center}
    \end{minipage}\hfill
    \begin{minipage}[t]{.49\linewidth}
        \begin{center}
\includegraphics[width=.95\linewidth]{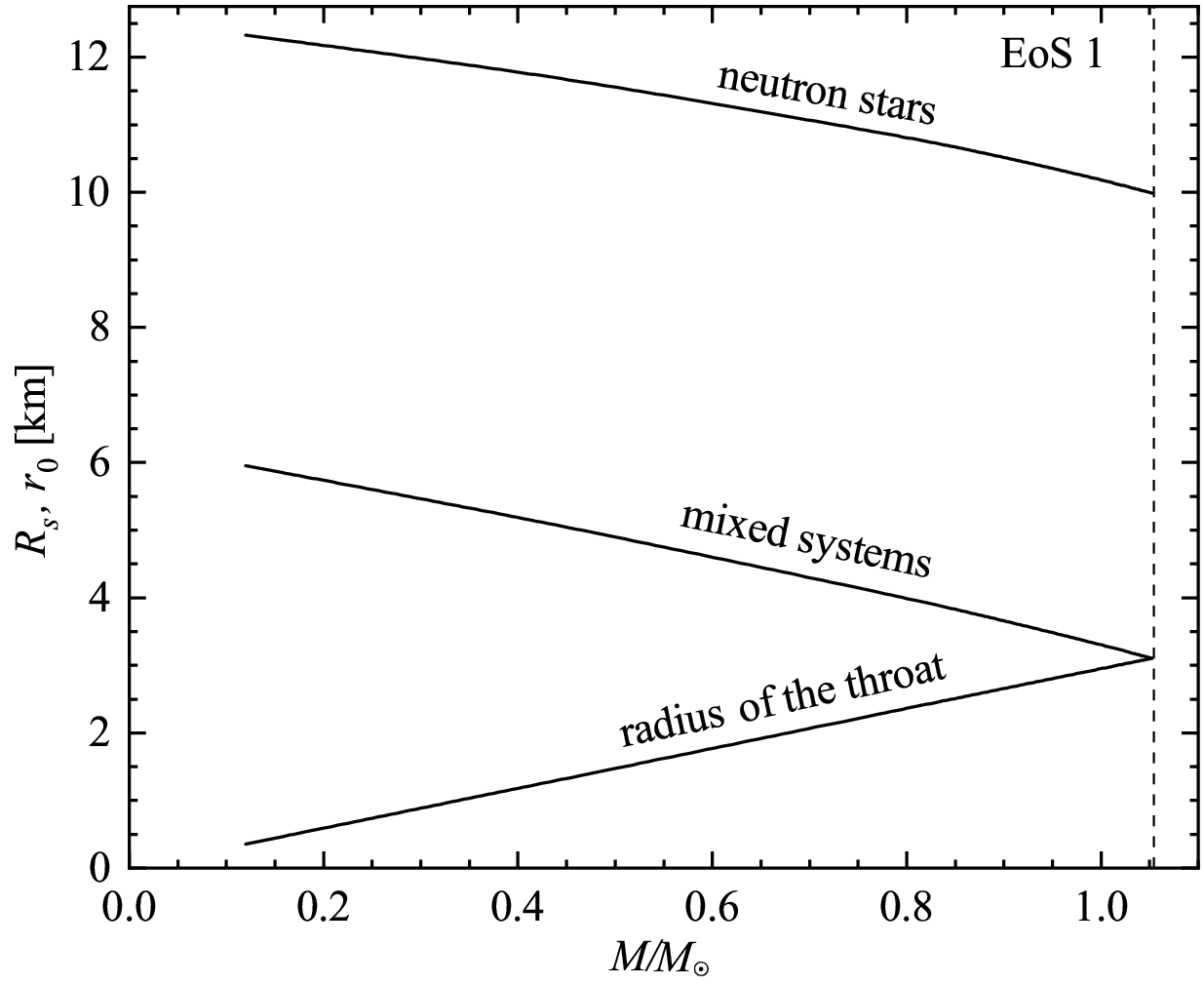}
\end{center}
    \end{minipage}
    \begin{minipage}[t]{.49\linewidth}
        \begin{center}
\includegraphics[width=1.\linewidth]{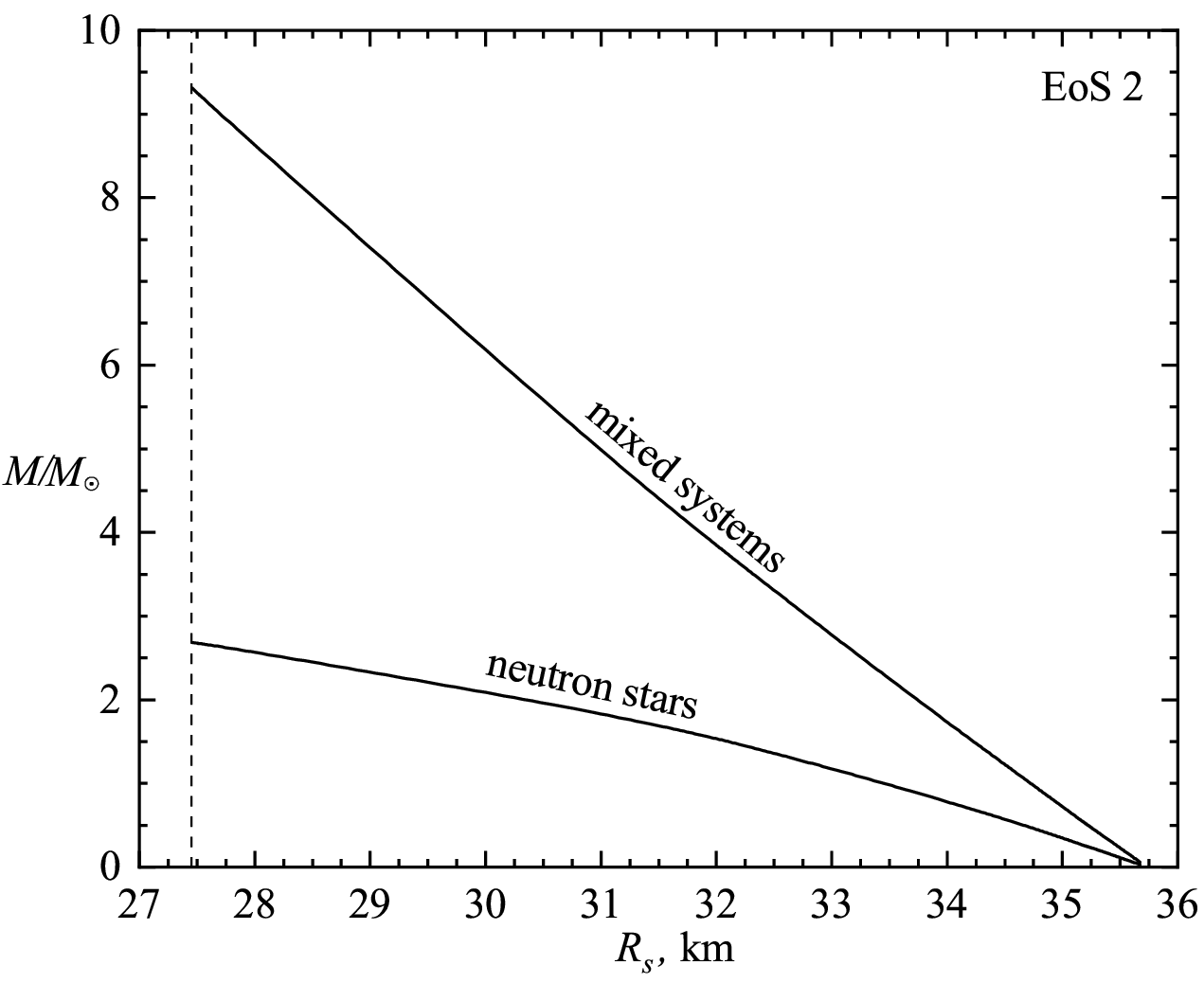}
        \end{center}
    \end{minipage}\hfill
    \begin{minipage}[t]{.49\linewidth}
        \begin{center}
\includegraphics[width=.98\linewidth]{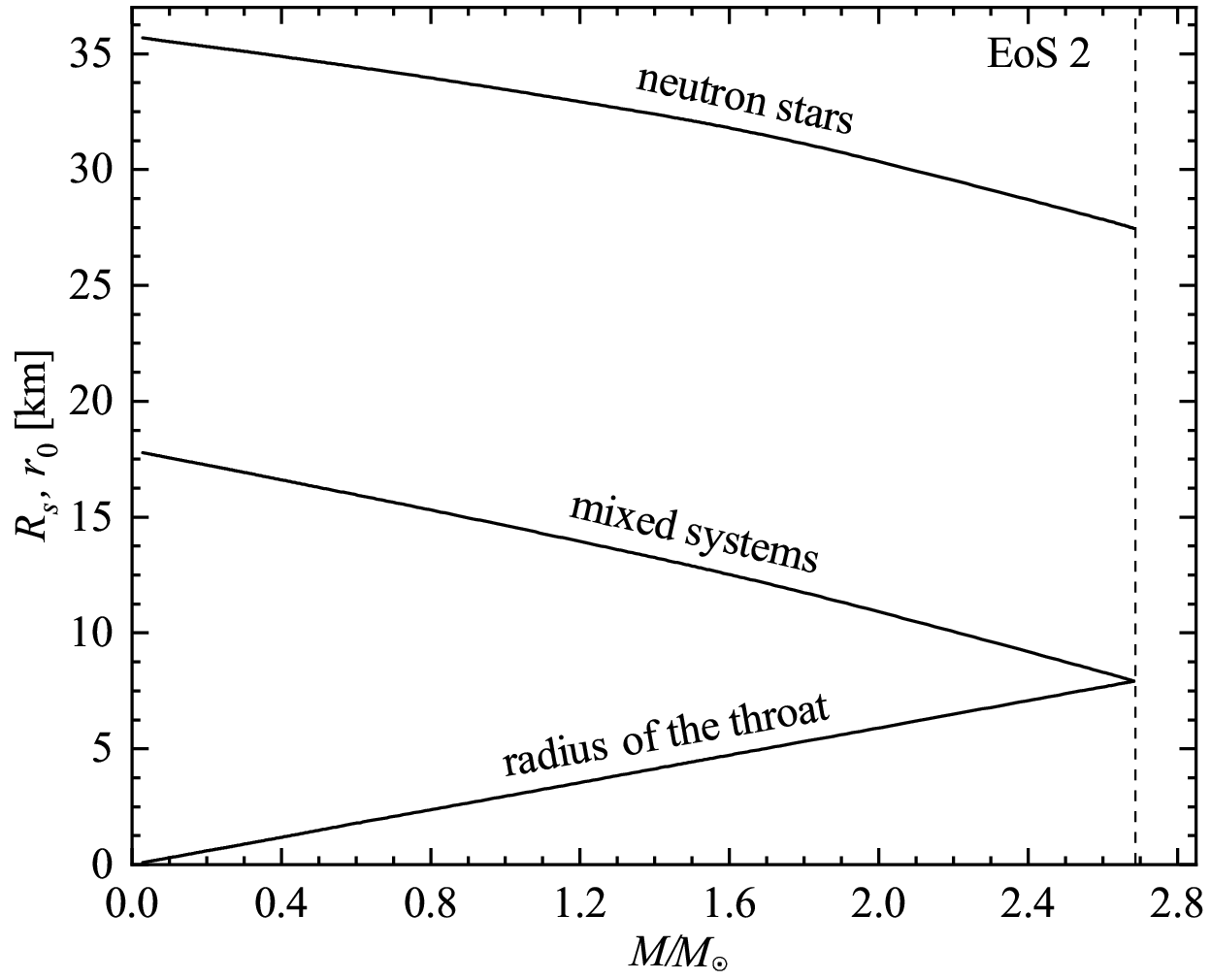}
\end{center}
    \end{minipage}
\caption{Left panels:
the ADM mass for ordinary neutron stars and mixed systems as a function of the radius. 
Right panels: the radius as a function of the ADM mass.
The vertical dashed lines correspond to the limiting 
    mixed system with $r_0\to R_s$.
}
\label{fig_Mass_radius}
\end{figure}

In this subsection, we construct mass-radius relations for the system under consideration and compare them with those of ordinary neutron stars 
with the same EoSs. To do this, it is necessary to choose appropriate values of the parameters appearing in the polytropic EoSs \eqref{polytrope} and \eqref{MassPolytropicEOS}.
For our purposes, we take for the EoS1 $n_1=1$ and $K_1=100 \, \text{km}^2 $ (in units $c=G=1$) \cite{Kokkotas:2000up}.
In turn, for the EoS2, we take the following set of parameters for the neutron fluid (in CGS units): 
$m_\mathrm{b}=1.66 \times 10^{-24}\, \text{g}$,
$n_\mathrm{b}^{(\mathrm{ch})} = 0.1\, \text{fm}^{-3}$,
$k=0.1$, and $n_2=1$ \cite{Salg1994}.

In general, the total mass of the system under consideration is given by \cite{Morris:1988cz, Morris:1988tu, Visser:1989kh, Visser:1989kg}
\begin{align}
\label{M_gen}
M(r) =\frac{r_0}{2 }+4\pi\int_{r_0}^{r}
\rho_\mathrm{tot}(r') r'^{2} dr'\, .
\end{align}
For the choice of the metric function $\lambda$ in the form \eqref{anz1WH},
 $\rho_\mathrm{tot}\equiv 0$ [see Eq.~\eqref{rho_tot}]. Then only the first term on the right-hand side of \eqref{M_gen} is present, which agrees with \eqref{ADMmass}.

Usually, the mass-radius curves for ordinary neutron stars are constructed by varying the central density of the fluid $\rho_c$.
As a result, for every particular value of  $\rho_c$, one has some particular values of $M$ and $R_s$.

In the case of the mixed system ``neutron star plus wormhole'' under consideration, we have the mass-radius relation in the form \eqref{ADMmass}.
In order to compare it with the relations for ordinary neutron stars, we may proceed as follows:
we take some $\rho_c$ and assume
that the radius of the fluid $R_s$ is equal to the radius of an ordinary neutron star obtained for the same  $\rho_c$.  
The corresponding results of calculations are given in the left panels of Fig.~\ref{fig_Mass_radius} for both EoSs.

Alternatively, we can take some $\rho_c$ and assume
that the mass of the mixed system is equal to the mass of an ordinary neutron star obtained for the same  $\rho_c$. 
The corresponding results of calculations are given in the right panels of Fig.~\ref{fig_Mass_radius} for both EoSs.

According to the observational data (see, e.g., Ref.~\cite{Ozel:2010fw}), the typical values of the mass of ordinary neutrons stars lie in the range $M\sim (1-2)\, M_\odot$,
while they radii are of the order of $R\sim 8-12\, \text{km}$. As seen from Fig.~\ref{fig_Mass_radius}, for the mixed systems under consideration, such characteristics 
are accessible both for the EoS1 (see the top left panel) and for the EoS2 (see the bottom right panel).

\section{Wormhole inside the black hole}
\label{WormholeinBlackHole}

First of all, we should note that the standard Reissner-Nordstr\"{o}m spacetime can be regarded as a wormhole because the geometry connects many universes, 
as can be seen from the corresponding Penrose diagram of the spacetime. 

We now consider an object that describes a wormhole inside a black hole whose geometry differs from the Reissner-Nordstr\"{o}m spacetime; 
that is, we consider a wormhole inside the event horizon. 
Such a wormhole could have a timelike throat. 
Then the object falling into the black hole via the event horizon goes first through the wormhole, then through the horizon, and eventually appears in another universe, 
where the black hole may behave as a white hole. 

In this section, we neglect the contribution of the matter. 
As an example, let us consider the ansatz 
\begin{align}
\label{WHinBH1}
\e^{-2\lambda}=\frac{\left( r - r_\mathrm{horizon} \right) \left( r^2 - {r_0}^2 \right)}{r^3}\, , \quad 
\e^{2\nu}= 1 - \frac{r_\mathrm{horizon} }{r} \, .
\end{align}
Here we assume that the radius $r_0$ of the throat is smaller than the Schwarzschild radius $r_\mathrm{horizon}$, $r_0<r_\mathrm{horizon}$. 
The behavior for large $r$ tells us that the ADM mass $M$ is given by $M=\frac{r_\mathrm{horizon}}{2}$. 
Compared with Eq.~(\ref{throat1}),  here we have 
\begin{align}
\label{WHinBH1B}
\e^{2\nu_0} = 1 - \frac{r_\mathrm{horizon} }{r_0} 
\, , \quad \nu_1 = \frac{r_\mathrm{horizon} }{{r_0}^2} \, , \quad 
\e^{2\lambda}\sim \frac{{r_0}^3}{2r_0 \left( r_0 - r_\mathrm{horizon} \right) \left( r - r_0 \right)}\, .
\end{align}
Therefore the metric given by Eq.~(\ref{WHinBH1}) has a wormhole throat at $r=r_0$, although the region around the throat is timelike. 

In the limit $r_0\to 0$, the metric given by Eq.~(\ref{WHinBH1}) reduces to the one of the Schwarzschild spacetime: $\e^{-2\lambda},\, \e^{2\nu}\to 1 - \frac{r_0}{r}=1 - \frac{2M}{r}$. 
In turn, in this limit, the functions $A$, $B$, $C$, and $V$ from Eq.~(\ref{ABCV}) vanish. 
In fact, we find 
\begin{align}
\label{ABCV2BB}
A\to&\, \frac{\e^{2\nu}}{\kappa^2} \left[ \e^{-2\lambda}\left( \frac{\nu' + \lambda'}{r} + \nu'' + \left( \nu' - \lambda' \right) \nu' + \frac{\e^{2\lambda} - 1}{r^2}\right) \right] 
\nonumber \\
\to&\,  \frac{\e^{4\nu}}{\kappa^2} \left[ \frac{- \frac{2r_0}{r^3} \left( 1 - \frac{r_0}{r} \right) - \frac{{r_0}^2}{r^4}}{2\left( 1 - \frac{r_0}{r} \right)^2}
+ \frac{\frac{{r_0}^2}{r^4}}{2\left( 1 - \frac{r_0}{r} \right)^2} + \frac{\frac{r_0}{r^3}}{1 - \frac{r_0}{r}} \right] = 0 \, , \nonumber \\
B=&\, 0 \, , \nonumber \\
C\to&\, \frac{\e^{2\lambda}}{\kappa^2} \left[
 - \e^{-2\lambda}\left( - \frac{\nu' + \lambda'}{r} + \nu'' + \left( \nu' - \lambda' \right) \nu' + \frac{\e^{2\lambda} - 1}{r^2}\right) \right] \to \e^{-4\nu} A \to 0
\, , \nonumber \\
V\to& \frac{\e^{-2\lambda}}{\kappa^2} \left( \frac{\lambda' - \nu'}{r} + \frac{\e^{2\lambda} - 1}{r^2} \right) 
\to \frac{\e^{-2\lambda}}{\kappa^2} \left( - \frac{\frac{r_0}{r^3}}{1 - \frac{r_0}{r}} + \frac{\frac{r_0}{r^3}}{1 - \frac{r_0}{r}} \right) = 0 \, .
\end{align}
Therefore, in the limit $r_0\to 0$, the scalar fields $\phi$ and $\chi$ decouple and the model reduces to the vacuum Einstein gravity without a cosmological constant.

We now consider the orbit of a test particle to check the motion of the particle after it penetrates the horizon. 
The motion can be described by the corresponding geodesic equation obtained from 
the Lagrangian 
\begin{align}
\label{Lag1}
L= m \sqrt{ - g_{\mu\nu} \dot x^\mu (\tau) \dot x^\nu (\tau)} = m \sqrt{ \left( 1 - \frac{r_\mathrm{horizon} }{r} \right)\left( \frac{dt}{d\tau} \right)^2 
 - \frac{\left( r - r_\mathrm{horizon} \right) \left( r^2 - {r_0}^2 \right)}{r^3} \left( \frac{dr}{d\tau} \right)^2 }\, .
\end{align}
Here $\tau$ is an affine parameter which parametrizes the orbit of the test particle. 
For simplicity, we do not consider here the motion in the angular directions.  
The Euler-Lagrange equations derived from the Lagrangian (\ref{Lag1}) are as follows: 
\begin{align}
\label{Lag2}
0 =&\, \frac{d}{d\tau} \left[ \frac{ m \left( 1 - \frac{r_\mathrm{horizon} }{r} \right) \frac{dt}{d\tau} }
{\sqrt{ \left( 1 - \frac{r_\mathrm{horizon} }{r} \right)\left( \frac{dt}{d\tau} \right)^2 
 - \frac{\left( r - r_\mathrm{horizon} \right) \left( r^2 - {r_0}^2 \right)}{r^3} \left( \frac{dr}{d\tau} \right)^2 }} \right]\, , \\
\label{Lag3}
0 =&\, \frac{d}{d\tau} \left[ \frac{ \frac{m r_\mathrm{horizon} \left( r - r_\mathrm{horizon} \right) \left( r - r_0 \right)}{r^3} \frac{dr}{d\tau} }
{\sqrt{ \left( 1 - \frac{r_\mathrm{horizon} }{r} \right)\left( \frac{dt}{d\tau} \right)^2 
 - \frac{\left( r - r_\mathrm{horizon} \right) \left( r^2 - {r_0}^2 \right)}{r^3} \left( \frac{dr}{d\tau} \right)^2 }} \right] \nonumber \\
&\, - \frac{m \left[ \frac{r_\mathrm{horizon} }{r^2} \left( \frac{dt}{d\tau} \right)^2 
 - \left( - \frac{3 r_\mathrm{horizon} {r_0}^2}{r^4} + \frac{2 {r_0}^2}{r^3} 
+ \frac{r_\mathrm{horizon}}{r^2} \right) \left( \frac{dr}{d\tau} \right)^2 \right] }
{2\sqrt{ \left( 1 - \frac{r_\mathrm{horizon} }{r} \right)\left( \frac{dt}{d\tau} \right)^2 
 - \frac{\left( r - r_\mathrm{horizon} \right) \left( r^2 - {r_0}^2 \right)}{r^3} \left( \frac{dr}{d\tau} \right)^2 }} \, .
\end{align}
The equation~(\ref{Lag2}) yields the conserved energy of the particle $E$, 
\begin{align}
\label{Lag4}
\frac{m \left( 1 - \frac{r_\mathrm{horizon} }{r} \right) \frac{dt}{d\tau} }
{\sqrt{ \left( 1 - \frac{r_\mathrm{horizon} }{r} \right)\left( \frac{dt}{d\tau} \right)^2 
 - \frac{\left( r - r_\mathrm{horizon} \right) \left( r^2 - {r_0}^2 \right)}{r^3} \left( \frac{dr}{d\tau} \right)^2 }} = E \, .
\end{align}

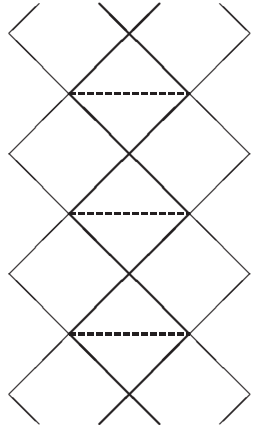
\begin{figure}[h]
\begin{center}

\unitlength=0.4mm
\begin{picture}(100,140)

\thicklines

\put(40,0){\line(1,1){30}}
\put(70,30){\line(-1,1){40}}
\put(30,70){\line(1,1){40}}
\put(70, 110){\line(-1,1){30}}

\put(60,0){\line(-1,1){30}}
\put(30,30){\line(1,1){40}}
\put(70,70){\line(-1,1){40}}
\put(30,110){\line(1,1){30}}

\multiput(30,30)(3, 0){13}{\line(1,0){2}}
\put(69, 30){\line(1,0){1}}
\multiput(30,70)(3, 0){13}{\line(1,0){2}}
\put(69, 70){\line(1,0){1}}
\multiput(30,110)(3, 0){13}{\line(1,0){2}}
\put(69, 110){\line(1,0){1}}

\thinlines

\put(20,0){\line(-1,1){10}}
\put(10,10){\line(1,1){20}}
\put(30,30){\line(-1,1){20}}
\put(10,50){\line(1,1){20}}
\put(30,70){\line(-1,1){20}}
\put(10,90){\line(1,1){20}}
\put(30,110){\line(-1,1){20}}
\put(10,130){\line(1,1){10}}

\put(80,0){\line(1,1){10}}
\put(90,10){\line(-1,1){20}}
\put(70,30){\line(1,1){20}}
\put(90,50){\line(-1,1){20}}
\put(70,70){\line(1,1){20}}
\put(90,90){\line(-1,1){20}}
\put(70,110){\line(1,1){20}}
\put(90,130){\line(-1,1){10}}

\end{picture}

\caption{The Penrose diagram of wormholes inside black holes.
The thick lines correspond to horizons, the thin lines~$-$ to infinities, and the dashed lines~-\,-\,- to the throat $r=r_0$. 
}\label{FIG1}
\end{center}
\end{figure}

We now choose $\tau$ to be the proper time of the test particle, which is defined by 
\begin{align}
\label{Lag5}
- d\tau^2 = g_{\mu\nu} dx^\mu dx^\nu = - \left( 1 - \frac{r_\mathrm{horizon} }{r} \right) dt^2 
+ \frac{\left( r - r_\mathrm{horizon} \right) \left( r^2 - {r_0}^2 \right)}{r^3} dr^2 \, . 
\end{align}
This gives 
\begin{align}
\label{Lag6}
\left( \frac{dt}{d\tau} \right)^2 = \frac{r^2 - {r_0}^2}{r^2} \left( \frac{dr}{d\tau} \right)^2 + \frac{r}{r - r_\mathrm{horizon}} \, .
\end{align}
Then Eq.~(\ref{Lag4}) is simplified to be 
\begin{align}
\label{Lag7}
m \left( 1 - \frac{r_\mathrm{horizon} }{r} \right) \frac{dt}{d\tau} = E \, .
\end{align}
Expressing $\frac{dt}{d\tau}$ from  Eq.~(\ref{Lag7}) and substituting it into Eq.~(\ref{Lag6}), we find 
\begin{align}
\label{Lag8}
\frac{r^2 - {r_0}^2}{r^2} \left( \frac{dr}{d\tau} \right)^2 = \frac{E^2 r^2}{m^2 \left( r -  r_\mathrm{horizon} \right)^2} - \frac{r}{r - r_\mathrm{horizon}} 
= \frac{E^2 r^2}{m^2 \left( r -  r_\mathrm{horizon} \right)^2} \left[ 1 - \frac{m^2}{E^2} \left( 1 - \frac{r_\mathrm{horizon}}{r} \right) \right] \, ,
\end{align}
which is consistent with Eq.~(\ref{Lag3}). 
We should note that $\left[1 - \frac{m^2}{E^2} \left( 1 - \frac{r_\mathrm{horizon}}{r} \right)\right]>0$, as long as $\frac{m^2}{E^2}<1$, as usually assumed. 
Therefore, $\frac{dr}{d\tau}$ does not vanish as long as $r>r_0$. 
Thus, if $\frac{dr}{d\tau}<0$ at a certain instant of time, we find $\frac{dr}{d\tau}<0$ 
until the particle reaches the throat as $r\to r_0$. 
Furthermore, when $r\gtrsim r_0$, Eq.~(\ref{Lag8}) has the following form: 
\begin{align}
\label{Lag9}
\left( r - r_0 \right) \left( \frac{dr}{d\tau} \right)^2 \sim \left( \frac{2 C}{3} \right)^2 
\equiv \frac{E^2 {r_0}^2}{m^2 \left( r_0 -  r_\mathrm{horizon} \right)^2} \left[ 1 - \frac{m^2}{E^2} \left( 1 - \frac{r_\mathrm{horizon}}{r_0} \right) \right] \, ,
\end{align}
which gives 
\begin{align}
\label{Lag10}
\left( r - r_0 \right)^\frac{3}{2} \sim C \left(\tau_0 - \tau\right)\quad \textrm{or} \quad 
r\sim r_0 + C^\frac{2}{3} \left(\tau_0 - \tau\right)^\frac{2}{3} \, .
\end{align}
Therefore, the test particle reaches the throat in a finite proper time and after that, it appears in another universe and goes through the horizon. 
Therefore, instead of a black hole, there is a white hole in another universe. 
We may consider the inverse process, that is, the particle falling into the black hole in another universe may appear in our universe from a white hole. 
The wormholes inside the black holes connect an infinite number of universes, as in the Reissner-Nordstr\"{o}m black hole, although the object considered in this paper has only 
event (outer) horizons. 
In Fig.~\ref{FIG1}, the Penrose diagram of the wormholes inside the black holes is given. 
This Penrose diagram has been also found in Ref.~\cite{Simpson:2018tsi}.

We now consider the Hawking temperature of the geometry given by Eq.~(\ref{WHinBH1}). 
Near the horizon $r\sim r_\mathrm{horizon}$, the metric has the following form: 
\begin{align}
\label{Htemp1}
ds^2 \sim -  \frac{r - r_\mathrm{horizon}}{r_\mathrm{horizon}} dt^2 
+ \frac{{r_\mathrm{horizon}}^3}{\left( r - r_\mathrm{horizon} \right) \left( {r_\mathrm{horizon}}^2 - {r_0}^2 \right)} dr^2 
+ r^2 \left( d\vartheta^2 + \sin^2\vartheta d\varphi^2 \right)\, .
\end{align}
We Wick-rotate the time coordinate $t$ by changing $t=\imath p$ and define a new variable $q$ as 
\begin{align}
\label{Htemp2}
q \equiv 2 \sqrt{ \frac{{r_\mathrm{horizon}}^3 \left( r - r_\mathrm{horizon} \right) }{{r_\mathrm{horizon}}^2 - {r_0}^2}} \, ,
\end{align}
using which we rewrite the metric (\ref{Htemp1}) in the form 
\begin{align}
\label{Htemp3}
ds^2 \sim \frac{{r_\mathrm{horizon}}^2 - {r_0}^2}{4{r_\mathrm{horizon}}^4} q^2 dp^2 + dq^2 + {r(q)}^2 \left( d\vartheta^2 + \sin^2\vartheta d\varphi^2 \right)\, .
\end{align}
In order to avoid the conical singularity at $q=0$, the Euclidean time $p$ should have a periodicity 
$p\sim p + \frac{4\pi {r_\mathrm{horizon}}^2}{\sqrt{{r_\mathrm{horizon}}^2 - {r_0}^2}}$, which gives the Hawking temperature $T_\mathrm{H}$, 
\begin{align}
\label{Htemp4}
T_\mathrm{H} = \frac{\sqrt{{r_\mathrm{horizon}}^2 - {r_0}^2}}{4\pi {r_\mathrm{horizon}}^2} 
= \frac{\sqrt{4M^2 - {r_0}^2}}{16\pi M^2} \, .
\end{align}
In the case of the standard Schwarzschild black hole, which corresponds to the limit $r_0 \to 0$, we find the usual result, $T_\mathrm{H}\to \frac{1}{8\pi M}$. 
Therefore, Eq.~(\ref{Htemp4}) tells us that the temperature of the black hole with a wormhole considered in this paper is lower than that of 
the standard Schwarzschild black hole with the identical ADM mass $M$. 
Especially, in the limit $r_0\to r_\mathrm{horizon}$, the Hawking temperature vanishes,  $T_\mathrm{H}\to 0$, which is similar to the extremal limit 
of the Reissner-Nordstr\"{o}m black hole. 

We may assume that the thermodynamical entropy $\mathcal{S}$ is given by the area of the horizon, 
\begin{align}
\label{ent1}
\mathcal{S}=4\pi {r_\mathrm{horizon}}^2\, .
\end{align}
Then Eq.~(\ref{Htemp4}) yields 
\begin{align}
\label{Htemp5}
T_\mathrm{H} = \frac{\sqrt{\frac{\mathcal{S}}{4\pi} - {r_0}^2}}{\mathcal{S}} \, ,
\end{align}
which gives 
\begin{align}
\label{Htemp6}
\frac{d\mathcal{S}}{d T_\mathrm{H}} = \frac{8\pi \mathcal{S}^2 \sqrt{\frac{\mathcal{S}}{4\pi} - {r_0}^2}}{8\pi {r_0}^2 - \mathcal{S}} \, .
\end{align}
By using the thermodynamic identity, the specific heat $C$ is given by 
\begin{align}
\label{ent2}
C \equiv \frac{dQ}{dT_\mathrm{H}}= T_\mathrm{H} \frac{d\mathcal{S}}{d T_\mathrm{H}}
= - \frac{2 \mathcal{S} \left(\mathcal{S} - 4 \pi {r_0}^2\right)}{\mathcal{S} - 8\pi {r_0}^2} 
= - \frac{8 \pi {r_\mathrm{horizon}}^2 \left({r_\mathrm{horizon}}^2 - {r_0}^2\right)}{{r_\mathrm{horizon}}^2 - 2 {r_0}^2} \, ,
\end{align}
where $Q$ is the heat. 
For the Schwarzschild black hole, i.e., when $r_0\to 0$, we find $C\to - 8\pi {r_\mathrm{horizon}}^2 = - 32 \pi M^2$, which is always negative.  
The specific heat $C$ in Eq.~(\ref{ent2}) is also negative when the radius of the black hole is large, $r_\mathrm{horizon} \gg r_0$, as for the standard Schwarzschild black hole. 
In turn, the specific heat $C$ vanishes in the extremal limit $r_\mathrm{horizon} \to r_0$. 
One can also see that the specific heat diverges when $r_\mathrm{horizon} = \sqrt{2}\, r_0$, 
which implies that there is a maximum $T_\mathrm{H}^\mathrm{max}$ 
in the Hawking temperature $T_\mathrm{H}$, which is also clear from Eq.~(\ref{Htemp4}), 
\begin{align}
\label{ent3}
T_\mathrm{H}^\mathrm{max} = \frac{1}{8\pi r_0}\, .
\end{align}

The above properties are very similar to the properties of the Reissner-Nordstr\"{o}m black hole, whose specific heat and temperature vanish 
in the extremal limit where the radius of the outer horizon coincides with the radius of the inner horizon, 
and there is a maximum in the Hawking temperature in the Reissner-Nordstr\"{o}m black hole, where the specific heat diverges. 
In our model, the radius $r_0$ of the throat plays the role of the radius of the inner horizon in the Reissner-Nordstr\"{o}m black hole. 

\section{Summary and Discussion}
\label{SummaryDiscussion}

In this paper, we consider two exotic objects:
(i)~a wormhole whose throat is hidden by a stellar object like a neutron star; 
and (ii)~a wormhole inside a black hole. 
These objects can be realized within the models which include two scalar fields by using the formulation of Ref.~\cite{Nojiri:2020blr}. 
Within the formulation, we can construct models that realize an arbitrarily given spherically symmetric and static/time-dependent geometry. 
In the original formulation of Ref.~\cite{Nojiri:2020blr}, however, the models include ghosts, which make the model physically inconsistent 
both as classical theory and as quantum theory. 
In the works \cite{Nojiri:2023dvf, Nojiri:2023zlp, Elizalde:2023rds, Nojiri:2023ztz}, however, it has been shown that the ghosts can be eliminated 
by the corresponding constraints. 
The latter is given by adding terms including the Lagrange multiplier fields to the action. 
These constraints are similar to the mimetic constraint of Ref.~\cite{Chamseddine:2013kea}. 
In the original mimetic model, non-dynamical dark matter effectively appears. 
Our model could be regarded as an extension of the mimetic theory and we realized effectively more general but non-dynamical matter, 
including exotic matter like a phantom. 
We also compared the mass-radius relations of the mixed configurations ``neutron star plus wormhole'' 
with those of ordinary neutron stars and demonstrated that it is possible to get masses and sizes of the mixed systems comparable to those typical of neutron stars. 

We have also investigated the structure of the exotic objects. 
For the wormhole whose throat is hidden inside the star, the ADM mass is proportional to the radius of the star, as in the case of the Schwarzschild black hole, 
although the mass of ordinary stars constructed with some \textit{ad hoc} profile of matter similar to that of mixed systems 
is proportional to the third power of the radius. 
We have also found that the null energy condition and therefore all the energy conditions are violated near the throat.  
The violation does not, however, generate any instability because the effective fluid is given by the non-dynamical scalar fields. 
The scalar fields do not propagate and do not fluctuate, and therefore there does not appear sound generated by the oscillation of the effective fluid. 
In this sense, the effective fluid is frozen. 

For the wormhole, the throat is timelike and we have clarified the causal structure. 
This wormhole has properties similar to those of the Reissner-Nordstr\"{o}m black hole, which has an electric charge and two horizons. 
Both in the case of the Reissner-Nordstr\"{o}m black hole and for the wormhole inside the black hole, an infinite number of universes are connected via horizons and/or throats. 
Furthermore, in the case of the Reissner-Nordstr\"{o}m black hole, there appears an extremal limit where the radii of the two horizons coincide with each other, 
and, in the limit, the Hawking temperature vanishes. 
In the case of the wormhole inside the black hole considered in this paper, the extremal limit appears when the radius of the throat coincides with the radius of the horizon. 
In this limit, the Hawking temperature vanishes, again. 

We may speculate how one can describe the creation of such exotic objects. 
As is clear from the arguments given in the subsection~\ref{sec:Reconstruction}, we may consider the time evolution of the spacetime. 
For this case, by starting from some appropriate initial configuration of the spacetime, of the scalar fields, and of matter, 
we may describe the formation of exotic objects. 
For example, we may start with an almost flat background and a spherically symmetric distribution of low-density matter. 
Alternatively, we may consider as the initial condition that there are a wormhole and spherically symmetric low-density matter. 
Then, for the time-dependent and spherically symmetric spacetime, instead of (\ref{FRN2}), the conservation law yields 
\begin{align}
\label{cnsvr1}
0 = - \dot\rho - \dot\lambda \left( \rho + p \right) \, , \quad 
0 = \nu' \left( \rho + p \right) + p' \, .
\end{align}
In the time-dependent case, the above equations should be generalized as follows:
\begin{align}
\label{conserv1}
\frac{\partial T_t^t}{\partial t}+\frac{\partial T_t^r}{\partial r}+\left(T_t^t-T_r^r\right)\frac{\partial \lambda}{\partial t}
+T^r_t\left[\frac{\partial }{\partial r}\left(\nu+\lambda\right)+\frac{2}{r}\right]=0 \, ,
\end{align}
and
\begin{align}
\label{conserv2}
\frac{\partial T_r^t}{\partial t}+\frac{\partial T_r^r}{\partial r}+T_r^t\frac{\partial }{\partial t}\left(\nu+\lambda\right) 
+ \left(T_r^r-T_t^t\right)\frac{\partial \nu}{\partial r}+\frac{2}{r}\left(T^r_r-T_\vartheta^\vartheta\right)=0\, ,
\end{align}
respectively. 
Since in general $T_r^t\neq 0$, the equations \eqref{cnsvr1} are, strictly speaking, incorrect. 
By analogy with Eq.~(\ref{FRN3}), we can integrate the above equations as follows: 
\begin{align}
\label{FRN3B}
\nu (t,r) = - \int^{p(t,r)}\frac{dp}{\rho(p) + p} + \nu_0 (t) \, , \quad 
\lambda (t,r) = - \int^{\rho (t,r)}\frac{d\rho}{\rho + p(\rho)} + \lambda_0 (r) \, . 
\end{align}
Here $\nu_0(t)$ and $\lambda_0(r)$ are arbitrary functions of $t$ and $r$, respectively. 
Therefore we may consider them as a proper initial distribution of the matter. 

It might be interesting to consider the fusion of two black holes or neutron stars. 
If one or two of the black holes or neutron stars have wormhole(s) inside, during the fusion, there might appear an energy flow between our universe and another universe. 
If the flux goes to another universe, the radiation after the fusion might become smaller. 
Conversely, if the flux comes from another universe, the radiation might be enhanced. 
Anyway, such fusions may provide us with some clue concerning 
the wormhole and another universe.

\section*{Acknowledgements}

This work was partially supported by MICINN (Spain), project
PID2019-104397GB-I00  and by the program Unidad de Excelencia
Maria de Maeztu CEX2020-001058-M, Spain (S.D.O). S.N. was partly
supported by MdM Core visiting professorship at ICE-CSIC,
Barcelona.
This research was funded by the Committee of Science of the Ministry of Science and Higher Education of the Republic of Kazakhstan (Grant No.~BR21881941). 

\appendix

\section{Deriving the field equations (\ref{I10}) from the Einstein equations (\ref{gb4bD4})}\label{Bianchi}

In this appendix, we show that the field equations (\ref{I10}) for the scalar fields $\phi$ and $\chi$  can be obtained from 
the Einstein equations (\ref{gb4bD4}) by using the Bianchi identity and the conservation law. 

Multiplying Eqs.~(\ref{gb4bD4}) by $\nabla^\nu$ and using the Bianchi identity $\nabla^\nu \left(- R_{\mu\nu} + \frac{1}{2} g_{\mu\nu} R\right) =0$ 
and the conservation law $\nabla^\nu T_{\mathrm{matter}\, \mu\nu} =0$, we obtain, 
\begin{align}
\label{gb4bD4AA2}
0=&\, \frac{1}{2} \partial_\mu \left\{
 - \frac{1}{2} A (\phi,\chi) \partial_\rho \phi \partial^\rho \phi
 - B (\phi,\chi) \partial_\rho \phi \partial^\rho \chi
 - \frac{1}{2} C (\phi,\chi) \partial_\rho \chi \partial^\rho \chi - V (\phi,\chi)\right\} \nonumber \\
&\, + \frac{1}{2} \nabla^\nu \left\{ A (\phi,\chi) \partial_\mu \phi \partial_\nu \phi
+ B (\phi,\chi) \left( \partial_\mu \phi \partial_\nu \chi
+ \partial_\nu \phi \partial_\mu \chi \right)
+ C (\phi,\chi) \partial_\mu \chi \partial_\nu \chi \right\} \nonumber \\
=&\, \frac{1}{2} \left\{
 - \frac{1}{2} \left(A_\phi \partial_\mu \phi + A_\chi \partial_\mu \chi \right) \partial_\rho \phi \partial^\rho \phi 
 - A \partial^\rho \phi \nabla_\mu \partial_\rho \phi 
 - \left(B_\phi \partial_\mu \phi + B_\chi \partial_\mu \chi \right) \partial_\rho \phi \partial^\rho \chi
 - B \left( \nabla_\mu \partial_\rho \phi \partial^\rho \chi + \partial_\rho \phi \nabla_\mu \partial^\rho \chi \right) \right. \nonumber \\
&\, \left. - \frac{1}{2} \left(C_\phi \partial_\mu \phi + C_\chi \partial_\mu \chi \right) \partial_\rho \chi \partial^\rho \chi 
 - C \partial_\rho \chi \nabla_ \mu \partial^\rho \chi 
 - V_\phi \partial_\mu \phi - V_\chi \partial_\mu \chi \right\} \nonumber \\
&\, + \frac{1}{2} \left\{ 
 \left(A_\phi \partial^\nu \phi + A_\chi \partial^\nu \chi \right) \partial_\mu \phi \partial_\nu \phi
+ A \left( \nabla^\nu \partial_\mu \phi \partial_\nu \phi + \partial_\mu \phi \nabla^\nu \partial_\nu \phi \right)\right. \nonumber \\
&\, 
+ \left(B_\phi \partial^\nu \phi + B_\chi \partial^\nu \chi \right) \left( \partial_\mu \phi \partial_\nu \chi + \partial_\nu \phi \partial_\mu \chi \right) 
+ B \left( \nabla^\nu \partial_\mu \phi \partial_\nu \chi + \partial_\mu \phi \nabla^\nu \partial_\nu \chi 
+ \nabla^\nu \partial_\nu \phi \partial_\mu \chi + \partial_\nu \phi \nabla^\nu \partial_\mu \chi \right) \nonumber \\
&\, \left. + \left(C_\phi \partial^\nu \phi + C_\chi \partial^\nu \chi \right) \partial_\mu \chi \partial_\nu \chi 
+ C \left( \nabla^\nu \partial_\mu \chi \partial_\nu \chi + \partial_\mu \chi \nabla^\nu \partial_\nu \chi \right) \right\} \, .
\end{align}
Taking into account our assumptions (\ref{GBiv_time}) and (\ref{TSBH1}), the $\mu=t$ and $\mu=r$ components become 
\begin{align}
\label{gb4bD4AA3}
0=&\, \frac{1}{2} \left\{
\frac{1}{2} A_\phi \e^{-2\nu} - A \e^{-2\nu} \dot \nu - B \left( - \e^{-2\lambda} \nu' + \e^{-2\lambda} \nu' \right) 
 - \frac{1}{2} C_\phi \e^{-2\lambda} + C \e^{-2\lambda} \dot\lambda - V_\phi \right\} \nonumber \\
&\, + \frac{1}{2} \left\{ - A_\phi \e^{-2\nu} 
+ A \left( \e^{-2\nu} \dot\nu + \e^{-2\nu} \dot\nu - \e^{-2\nu} \dot\lambda \right)
+ B_\chi \e^{-2\lambda} \right. \nonumber \\
&\, \left. + B \left( - \e^{-2\lambda} \nu' + \frac{2\e^{-2\lambda}}{r} + \e^{-2\lambda} \nu' - \e^{-2\lambda} \lambda' + \e^{-2\lambda} \nu' \right) 
 - C \e^{-2\lambda}\dot\lambda \right\} \nonumber \\
=&\, \frac{1}{2} \left\{ - \frac{1}{2} A_\phi \e^{-2\nu} + A \e^{-2\nu} \left( \dot\nu - \dot\lambda \right) 
+ B_\chi \e^{-2\lambda} - \frac{1}{2} C_\phi \e^{-2\lambda} + B \left( \frac{2\e^{-2\lambda}}{r} 
+ \e^{-2\lambda} \nu' - \e^{-2\lambda} \lambda' \right) - V_\phi \right\}  \, , \nonumber \\
0 =&\, \frac{1}{2} \left\{
\frac{1}{2} A_\chi \e^{-2\nu} 
 - A \e^{-2\nu} \nu'  - B \left( - \e^{-2\nu} \dot\lambda + \e^{-2\nu} \dot\lambda \right) 
 - \frac{1}{2} C_\chi \e^{-2\lambda} + C \e^{-2\lambda} \lambda' - V_\chi \right\} \nonumber \\
&\, + \frac{1}{2} \left\{ A \e^{-2\nu} \nu'  - B_\phi \e^{-2\nu} 
+ B \left( - \e^{-2\nu} \dot\lambda + \e^{-2\nu}\dot\nu - \e^{-2\nu} \dot\lambda 
+ \e^{-2\nu} \dot\lambda \right) 
+ C_\chi \e^{-2\lambda} \right. \nonumber \\
&\, \left. + C \left( - \e^{-2\lambda} \lambda' + \frac{2\e^{-2\lambda}}{r} + \e^{-2\lambda} \nu'  - \e^{-2\lambda} \lambda' \right) \right\} 
\nonumber \\
=&\, \frac{1}{2} \left\{ \frac{1}{2} A_\chi \e^{-2\nu} - B_\phi \e^{-2\nu} + B \left( \e^{-2\nu}\dot\nu - \e^{-2\nu} \dot\lambda \right) 
+ \frac{1}{2} C_\chi \e^{-2\lambda} + C \left( \frac{2\e^{-2\lambda}}{r} + \e^{-2\lambda} \nu'  - \e^{-2\lambda} \lambda' \right) - V_\chi \right\} \, ,
\end{align}
while other components vanish trivially. 
Here we have used the Christoffel symbols 
\begin{align}
\label{GBv0}
&\Gamma^t_{tt}=\dot\nu \, , \quad \Gamma^r_{tt} = \e^{-2(\lambda - \nu)}\nu' \, , \quad \Gamma^t_{tr}=\Gamma^t_{rt}=\nu'\, , \quad
\Gamma^t_{rr} = \e^{2\lambda - 2\nu}\dot\lambda \, , \quad \Gamma^r_{tr} = \Gamma^r_{rt} = \dot\lambda \, , \nonumber \\
& \Gamma^r_{rr}=\lambda'\, , \quad \Gamma^i_{jk} = \bar{\Gamma} ^i_{jk}\, ,\quad \Gamma^r_{ij}=-\e^{-2\lambda}r \bar{g}_{ij} \, , \quad
\Gamma^i_{rj}=\Gamma^i_{jr}=\frac{1}{r} \, \delta^i_{\ j}\,.
\end{align}
On the other hand, the field equations (\ref{I10})  for the scalar fields $\phi$ and $\chi$ have the following form: 
\begin{align}
\label{I10B}
0 =&\, - \frac{1}{2} A_\phi \e^{-2\nu} + A \e^{-2\nu} \left( \dot\nu - \dot\lambda \right) 
+ B_\chi \e^{-2\lambda} - \frac{1}{2} C_\phi \e^{-2\lambda} + B \e^{-2\lambda} \left( \frac{2}{r} + \nu' - \lambda' \right) - V_\phi   \, , \nonumber \\
0 =&\, \frac{1}{2} A_\chi \e^{-2\nu} - B_\phi \e^{-2\nu} + B \e^{-2\nu} \left( \dot\nu - \dot\lambda \right) 
+ \frac{1}{2} C_\chi \e^{-2\lambda} + C \e^{-2\lambda} \left( \frac{2}{r} + \nu' - \lambda' \right) - V_\chi \, .
\end{align}
The above expressions completely coincide with the equations in (\ref{gb4bD4AA3}). 
Therefore the field equations (\ref{I10}) can be surely obtained from the Einstein equations~(\ref{gb4bD4}). 

Especially, in the static case, we have from Eq.~(\ref{ABCV}), 
\begin{align}
\label{ABCV2}
A=&\, \frac{\e^{2\nu}}{\kappa^2} \left[ \e^{-2\lambda}\left( \frac{\nu' + \lambda'}{r} + \nu'' + \left( \nu' - \lambda' \right) \nu' + \frac{\e^{2\lambda} - 1}{r^2}\right) \right]
 - \e^{2\nu} \left( \rho + p \right) \, , \nonumber \\
B=&\, 0 \, , \nonumber \\
C=&\, \frac{\e^{2\lambda}}{\kappa^2} \left[
 - \e^{-2\lambda}\left( - \frac{\nu' + \lambda'}{r} + \nu'' + \left( \nu' - \lambda' \right) \nu' + \frac{\e^{2\lambda} - 1}{r^2}\right) \right] 
\, , \nonumber \\
V=& \frac{\e^{-2\lambda}}{\kappa^2} \left( \frac{\lambda' - \nu'}{r} + \frac{\e^{2\lambda} - 1}{r^2} \right) - \frac{1}{2} \left( \rho - p \right) \, .
\end{align}
Then the first equation in~(\ref{I10B}) is trivially satisfied, and the second equation gives 
\begin{align}
\label{2ndgrav}
& \frac{1}{2} A_\chi \e^{-2\nu} - B_\phi \e^{-2\nu} + B \e^{-2\nu} \left( \dot\nu - \dot\lambda \right) 
+ \frac{1}{2} C_\chi \e^{-2\lambda} + C \e^{-2\lambda} \left( \frac{2}{r} + \nu' - \lambda' \right) - V_\chi \nonumber \\
=&\, - \nu' \left( \rho + p \right) - p' =0 \, ,
\end{align}
which is nothing but the conservation law (\ref{FRN2}).

\end{document}